\documentclass[twocolumn,aps,prl]{revtex4}
\usepackage[english]{babel}
\usepackage[dvipdf]{graphicx}
\usepackage[dvips]{epsfig}
\usepackage{amssymb}
\usepackage{amsmath}
\usepackage{amsbsy}
\usepackage{color}
\usepackage{dcolumn}

\usepackage[section]{placeins}
\usepackage{color}
\usepackage{color}
\definecolor {myc} {rgb} {0,0,0}  

\begin{document}

\title{Interplay between viscosity and elasticity in freely expanding liquid sheets}

\author{Srishti Arora}
\author{Christian Ligoure}
\email{christian.ligoure@umontpellier.fr}
\author{Laurence Ramos}
\email{laurence.ramos@umontpellier.fr}

\affiliation{
Laboratoire Charles Coulomb UMR 5221,\\
 CNRS, Universit\'{e} de Montpellier,
 F-34095, Montpellier, France\\
}

\date{\today}

\begin{abstract}
We investigate the dynamics of freely expanding liquid sheets prepared with fluids with different rheological properties, (i) viscous fluids with a zero-shear viscosity $\eta_0$ in the range $(1-1000) \, \mathrm{mPa.s}$ and (ii)  viscoelastic fluids whose linear viscoelastic behavior in the frequency range $(0.1-100)$ rad/s can be accounted for by a Maxwell fluid model, with characteristic   elastic modulus, $G_0$, relaxation time, $\tau$, and zero-shear viscosity, $\eta_0=G_0 \tau$, can be tuned over several orders of magnitude. The sheets are produced by impacting a  drop of fluid on a small cylindrical solid target. For viscoelastic fluids, we show that, when $\tau$ is shorter than the typical lifetime of the sheet ($\sim 10$ ms), the dynamics of the sheet is similar to that of Newtonian viscous liquids with equal zero-shear viscosity. In that case, for little viscous samples ($\eta_0 <\sim 30$ mPa.s), the maximal expansion of the sheet, $d_{\rm{max}}$, is independent of $\eta_0$, whereas for more viscous samples, $d_{\rm{max}}$ decreases as $\eta_0$ increases. We provide a simple model for the dependence of the maximal expansion of the sheet with the viscosity that accounts well for our experimental data. By contrast, when $\tau$ is longer than the typical lifetime of the sheet, the behavior drastically differs. The sheet expansion is strongly enhanced as compared to that of viscous samples with comparable zero-shear viscosity, but is heterogeneous with the occurrence of cracks, revealing the elastic nature of the viscoelastic fluid.

\end{abstract}

\maketitle

\section {Introduction}

Upon hitting a solid or liquid surface, a drop can splash, partially or fully rebound, or remain on the surface and spread~\cite{Yarin2006, Josserand2016}. The maximal extension reached by the impinging drop when it spreads on the surface is very important for many practical configurations, including spray coating, pesticide application and ink-jet printing,  as it will define the mark made by the drop on the surface. For practical reasons fluids much more viscous than pure water and also with more complex rheological properties than pure newtonian fluids may be used. Accordingly, Newtonian fluids with broad ranges of viscosity~\cite{German2009a, Scheller1995} and a large variety of complex fluids have been investigated so far, ranging from dilute polymer solutions (see the review in Ref.~\cite{Bertola2013} and the references therein), to model viscoelastic surfactant solutions~\cite{Cooper-White2002}, yield-stress fluids~\cite{Nigen2005, German2009b, Luu2009, Luu2013, Saidi2010}, colloidal suspensions~\cite{Nicolas2005} and blood~\cite{Laan2014}. The importance of wettability for a yield stress fluid made of microgels~\cite{Luu2013}, and of {\color {myc} the dynamics of the contact line}  for dilute polymer solutions~\cite{Bartolo2007, Bertola2015} have been clearly evidenced in the dynamics of non-newtonian drops. However, even for simpler fluids as low viscosity Newtonian liquids, providing  a theoretical prediction for the maximal spreading of a drop as a function of the surface tension, contact angle, impact velocity and viscosity is not an easy task~\cite{Laan2014}. In the case of more viscous Newtonian fluids, the main difficulty in order to reach a theoretical prediction is to account for the viscous dissipation of the drop when it spreads, as complex flows are involved. In this optics, several approaches have been proposed yielding reasonably good agreements with experimental data~\cite{Scheller1995,Mao1997,Bolleddula2010}.

Here, we are interested in a related configuration, initially designed to suppress viscous dissipation ~\cite{Rozhkov2002, Rozhkov2004}, where a drop impacts a small target of size comparable to that of the drop. Upon impact the drop freely expands in air, reaches a maximal extension and then retracts due to surface tension~\cite{Rozhkov2002, Rozhkov2004, Villermaux2011, Vernay2015a, Vernay2015b}. Hence, the dissipation processes occurring at the fluid/surface interface are reduced, if not suppressed, possibly facilitating their modeling and/or experimental investigations. We note that this configuration is the discrete version of the Savart experiment, where a liquid jet hits normally a flat solid disk, resulting in a stationary planar liquid sheet~\cite{Savart1833}. In this case, detailed investigations of the effect of the viscosity and viscoelasticity  of the materials forming the sheets are scarce. A noticeable exception is the study of liquids comprising small amounts of long soluble polymers that has a dramatic effect on the destabilization processes due to the long time relaxation of the polymer chains~\cite{Rozhkov2006, Jung2011, Rozhkov2015}. In this paper, we thoroughly investigate experimentally how the viscosity and viscoelasticity of fluids impact the expansion of liquid sheets. To do so, we use Newtonian fluids whose viscosity varies on a very large range, and self-assembled networks that behave as pure Maxwell fluids and whose elastic modulus and characteristic relaxation time can be finely tuned over very large ranges, allowing one to independently investigate the role of viscosity and elasticity in the dynamics of viscoelastic sheets.

The paper is organized as follows. We first present the main experimental techniques and the samples investigated. The next section is devoted to the experimental results. We describe and quantify how sheets made of viscous and viscoelastic materials expand. Finally, we provide in the last section a modeling for the dynamics of viscous sheets and discuss the results for viscoelastic samples.

\section {Materials and Methods}
\label{Sec:MM}

\subsection{Experimental techniques}

Rheology is used to investigate the sample viscoelasticity and viscosity. A strain-controlled rheometer (Ares from TA instrument) and a stress-controlled rheometer (MCR 502 from Anton-Paar) equipped with a Couette geometry are used. Temperature is fixed at $T =25^{\rm{o}}$C for all rheology measurements. All other experiments are conducted at room temperature.
Surface tension measurements are performed using a Wilhelmy plate tensiometer.

Thin sheets freely expanding in air are produced by impacting a single drop of fluid on a solid target of size comparable to that of the drop. The set-up, initially designed by Rozhkov \textit{et al.}~\cite{Rozhkov2004}, has been described elsewhere~\cite{Vernay2015a}. In brief, we use a hydrophilic cylindrical target of diameter $d_{\rm{t}}=6$ mm, slightly larger than the drop diameter $d_0=(3.6 \pm 0.1)$ mm. The liquid drop is injected from a syringe pump through a needle placed vertically above the target. The drop falls from a distance of $91.0$ cm yielding a velocity at impact of $v_0 \approx 4$ m/s. In the following, the origin of time is taken at the drop impact. The size of the falling droplets is dictated by the inner diameter of the syringe and the equilibrium surface tension of the samples. In order to maintain a constant droplet size, needles with different diameters are used to account for the various equilibrium surface tension of the samples. Time series are recorded after the impact of the drop using a high speed camera Phantom V7.3 ($800$ pixels $\times$ $600$ pixels, operated at $6700$ frames per second).
After the drop impact, a liquid sheet freely expands in air. The sheet is bounded by a thicker rim that destabilizes into ligaments, which subsequently disintegrate into drops. The sheet then retracts due to surface tension.

Image J software is used to compute the time-evolution of the sheet diameter. The contour of the sheet is determined by thresholding the images, and using Wand(Tracing) tool, the sheet area $A$ is measured, from which an effective diameter is deduced using the simple geometric relation $d = \sqrt{4A/\pi}$.

\subsection{Experimental samples}

 Self-assembled transient networks consist of reversibly cross-linked polymers in solution, forming spontaneously three-dimensional networks at thermodynamical equilibrium that can transiently transmit elastic stresses over macroscopic distances. They exhibit simple rheological properties and are even able to fracture~\cite{Ligoure2013}. We investigate here networks comprising surfactant micelles~\cite{Tixier2010}, respectively surfactant-stabilized oil droplets~\cite{Michel2000}, dispersed in brine ($0.5$ M, respectively  $0.2$ M Nacl), and reversibly bridged by telechelic polymers. Micelles are composed of a mixture of cetylpyridinium chloride (CpCl) and sodium salicylate (NaSal) with a NaSal/CpCl molar ratio of $0.1$. Microemulsions are composed of decane droplets of $6$ nm diameter, stabilized by a mixture of CpCl and octanol, with a molar ratio of octanol/CpCl of $0.65$. Telechelic polymers are made of a long water-soluble polyethylene oxide chain of molecular weight $35$ kg/mol, at the extremity of which are grafted hydrophobic stickers that are short carboxylic chains $\rm{C_{n}}$ with $n=12$ or $n=18$.
For the micelle-based samples, the mass fraction of the micelles is $\varphi=(m_{\rm{CpCl}} + m_{\rm{NaSal}})/m_{\rm{total}}$, and the amount of polymer is $\beta = m_{\rm{polymer}}/(m_{\rm{CpCl}} + m_{\rm{NaSal}})$. Here $m_{\rm{CpCl}}$, $m_{\rm{NaSal}}$, and $m_{\rm{polymer}}$ are respectively the mass of CpCl, NaSal and polymer, and $m_{\rm{total}}$ is the total mass of the sample. $\varphi$ is fixed at $10$ \%, and the amount of polymer $\beta$ is varied between $0$ and $50$\%. For the microemulsion-based samples, we fix the average number of telechelic stickers per oil droplet $r=4$, and vary the mass fraction of oil droplet $\phi= (m_{\rm{hydrophobe}} + m_{\rm{oil}})/m_{\rm{total}}$ between $0.5$\% and $10$\%. Here $m_{\rm{hydrophobe}}$ is the mass of the hydrophobic part of surfactants and telechelic polymer and $m_{\rm{oil}}$ is the mass of oil.
To enhance the contrast of the liquid sheet, a dye (erioglaucine, concentration $2.5$ g/l) is eventually added to the samples~\cite{Vernay2015a, Vernay2015b}.

Additional tests are also performed at room temperature with purely viscous Newtonian mixtures of glycerol and water. The samples comprise eventually surfactant, CpCl ($5.88$ mM) and NaCl ($0.5$ M). The water/glycerol composition varies between $22$ wt\% and $97.5$ wt\%, yielding viscosity between $1.8$ mPa.s and $713$ mPa.s. We have checked that the addition of  CpCl and NaCl does not modify the sample viscosity.
Without CpCl and NaCl, the equilibrium surface tension of the water/glycerol mixtures ranges between $72.5$ mN.m$^{-1}$ and $62.9$ mN.m$^{-1}$. In the presence of surfactant and brine, the equilibrium surface tension is measured to decrease to $(38.5 \pm 2.0)$ mN.m$^{-1}$.

\begin{figure}
\includegraphics[width=0.5\textwidth]{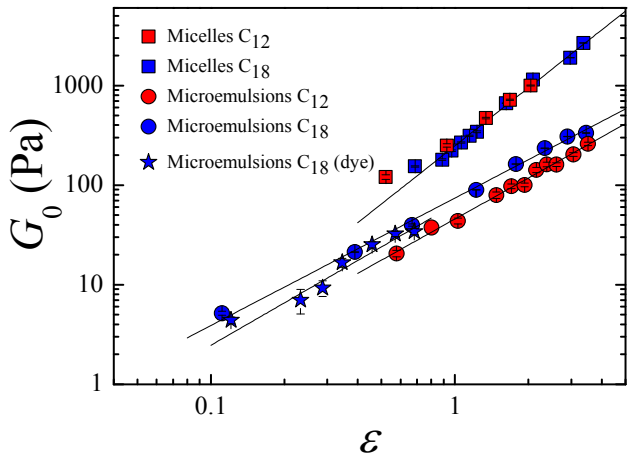}
\caption{(Color online) Elastic shear plateau modulus of viscoelastic samples  as a function of the distance from the percolation threshold. Symbols are data points and the lines are power law fits of the experimental data.}
\label{fig:elasticmodulus}
\end{figure}

\subsection{Linear viscoelasticity}

The sample composition is varied in order to tune the sample viscoelasticity. Our objective is to produce viscoelastic samples with relatively low zero shear viscosity so that drops can be produced and impacted on the target. As shown previously for similar systems~\cite{Michel2000, Tixier2010} the experimental samples investigated here (micelles and microemulsion reversibly linked by telechelic polymers) behave as purely viscous liquids below a percolation threshold and exhibit viscoelasticity above the threshold. In contrast to previous investigations~\cite{Tabuteau2009, Tabuteau2011, Tixier2010}, we here focus on the region of the phase-diagram close above the percolation threshold. Elastic samples above the percolation threshold behave as Maxwell fluids and are characterized by an elastic  shear plateau modulus, $G_0$, and a unique characteristic relaxation time, $\tau$. We have determined those parameters by fitting the frequency-dependence moduli of the samples with the Maxwell model (storage modulus
$G'(\omega)=\frac{G_0 (\omega \tau)^2}{1+ (\omega \tau)^2}$ and loss modulus  $G"(\omega)=\frac{G_0 \omega \tau}{1+ (\omega \tau)^2}$, with $\omega$ the pulsation). Note that for samples with relaxation times shorter than the inverse of the maximum frequency experimentally accessible ($\tau$ of the order of a few ms), we have cross-checked the numerical values extracted from the fit of the frequency dependence of the complex modulus with those obtained from the zero-shear viscosity measurements.
Here one varies the formulation, i.e. the mass fraction of oil droplets, $\phi$, for the emulsion-based samples, and the amount of telechelic polymers, $\beta$, for the micelle-based samples, in order to approach the percolation threshold where $G_0$ and $\tau$ vanish critically. One defines $\epsilon$ as the normalized distance from the percolation threshold: $\epsilon = \frac{\beta - \beta_c}{\beta_c}$ for the micelle-based samples and $\epsilon = \frac{\phi - \phi_c}{\phi_c}$ for the oil droplets-based samples.

\begin{figure}
\includegraphics[width=0.5\textwidth]{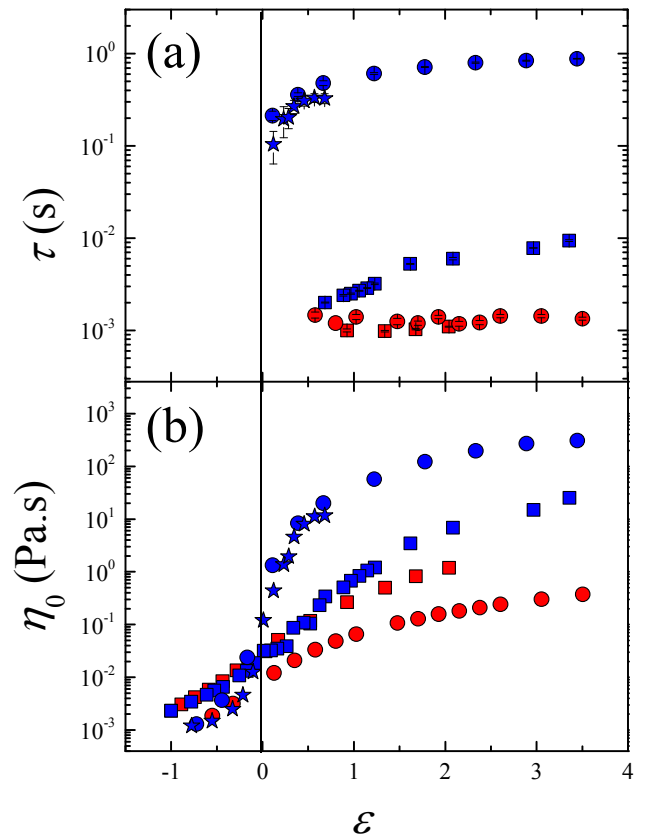}
\caption{(Color online) Characteristic relaxation time (a) and zero-shear viscosity (b) as a function of the distance from the percolation threshold. Symbols are the same as in fig.~\ref{fig:elasticmodulus}.}
\label{fig:tau}
\end{figure}

We show in figure~\ref{fig:elasticmodulus} the evolution of the shear  elastic plateau modulus with $\epsilon$ for the $4$ classes of samples (micelle and microemulsion comprising telechelic polymers with $\rm{C_{18}}$- and $\rm{C_{12}}$-long hydrophobic stickers). In all cases, close to the percolation threshold the evolution of the elastic modulus with the distance from the percolation threshold can be approximated by a power law fit: $G_0=A\epsilon^p$. Numerical values of the percolation thresholds, and of the exponent $p$, as derived from a power law fit of the data, are given in Table~\ref{tab:percolation}. Notably, we found for each class of samples (micelle-based and microemulsion-based networks) comparable  values for the percolation threshold and for the exponent, $p$, which are found independent of the telechelic polymer used.

\begin{table}[h!]
\begin{center}
 \begin{tabular}{||c c c c||}
 \hline
 Sample & $\phi_c (\%) $&$ \beta_c (\%)$& $p$ \\ [0.5ex]
 \hline\hline
 Micelles $ \mathrm{C_{18}}$ & & 13.5 & 1.76 \\
Micelles $\mathrm{C_{12}}$ &  & 11.5 & 1.76\\
 Microemulsion $\mathrm{C_{18}}$& 1.8 &  & 1.28\\
 Microemulsion $\mathrm{C_{12}} $& 2.2 &  & 1.37\\[1ex]
 \hline

\end{tabular}
\end{center}
 \caption{Percolation thresholds and critical exponents of the shear plateau moduli for the four classes of transient networks }
\label{tab:percolation}
\end{table}

On the other hand, the characteristic relaxation times $\tau$ strongly vary, as shown in figure~\ref{fig:tau}a, where the evolution of $\tau$  with $\epsilon$ is plotted for the $4$ classes of samples investigated. Indeed, changing the length of the hydrophobic carboxylic stickers leads to a change of the average residences time of the stickers in the micelles and oil droplets, implying in turn modifications of the characteristic viscoelastic relaxation times. We note that the relaxation times for the $\rm{C_{18}}$-microemulsion-based samples are systematically about two orders of magnitude larger than those for the $\rm{C_{18}}$-micelle-based samples.
Using the four classes of samples, one therefore has access to viscoelastic samples whose relaxation times span almost three orders of magnitude (from $1$ ms to $880$ ms). The zero-shear viscosity, $\eta_0$, of the samples, below and above the percolation threshold (above percolation, $\eta_0=G_0 \tau$) is plotted in figure~\ref{fig:tau}b. Below percolation ($\epsilon < 0$), the viscosity of the $\rm{C_{12}}$- and $\rm{C_{18}}$-based samples are equal, and weakly increases with $\epsilon$. Above percolation, $\eta_0$ increases more sharply with $\epsilon$, and, as expected, the viscosity of the $\rm{C_{12}}$- and the $\rm{C_{18}}$-based samples strongly differ. Overall the zero-shear viscosity spans more than $6$ orders of magnitude, from viscosity close to the one of water ($0.0012$ Pa.s) to $307$ Pa.s.
Notably, Maxwell fluids that display the same zero-shear viscosity but with drastically different characteristic relaxation time are available, allowing one to decouple the effect of viscosity and of elasticity in the processes at play in the dynamics of freely expanding sheets.

\section {Experimental results}
\label{Sec:results}

\subsection{Dynamics of viscous and viscoelastic sheets}

\subsubsection{Viscous samples}

\begin{figure}
\includegraphics[width=0.5\textwidth]{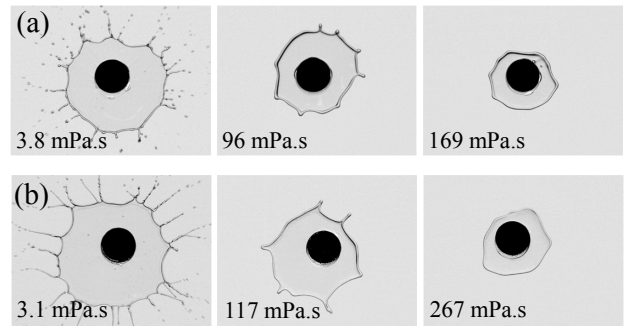}
\caption{Snapshots of (a) viscous sheets made of water/glycerol mixtures and (b) sheets made of $\rm{C_{12}}$-based micellar system, of various zero-shear viscosities as indicated. In (b), the sample with the lowest viscosity $\eta_0=3.1$ mPa.s is viscous and the two other samples are viscoelastic. Images are taken at the maximal expansion of the sheets.}
\label{fig:imagesViscous}
\end{figure}

\begin{figure}
\includegraphics[width=0.5\textwidth]{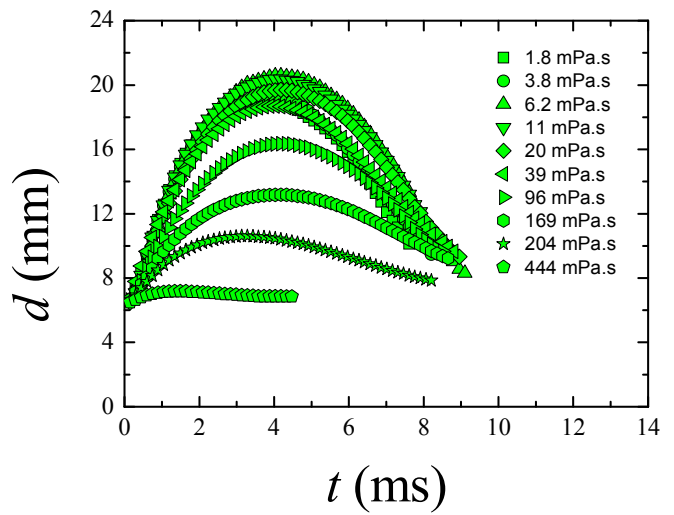}
\caption{(Color online) Time evolution of the sheet diameter for glycerol/water mixtures of various viscosities as indicated in the legend.}
\label{fig:dversust}
\end{figure}

\begin{figure}
\includegraphics[width=0.5\textwidth]{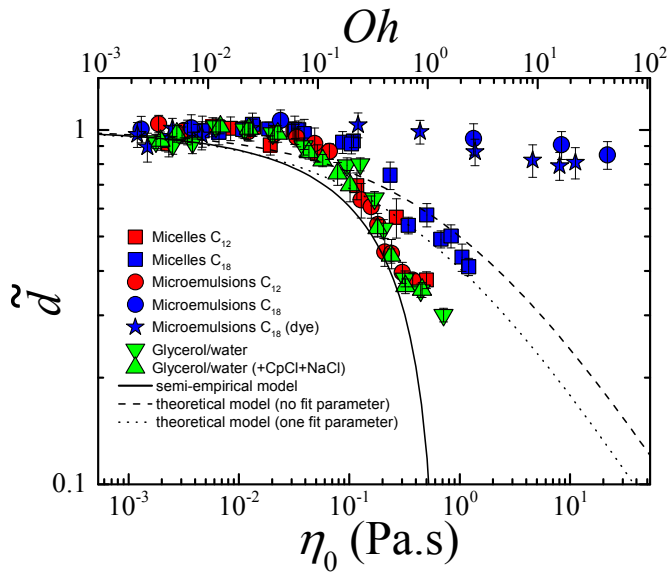}
\caption{(Color online) Maximum diameter (normalized by its value at low viscosity), as a function of zero-shear viscosity (bottom x axis) and Ohnesorge number (top x axis) for several classes of samples, as indicated in the legend. The symbols are data points and the lines are theoretical and semi-empirical modeling for viscous samples and viscoelastic samples for which $De \ll 1$ (see text).}
\label{fig:dmax}
\end{figure}

Figure~\ref{fig:imagesViscous} displays a series of images of sheets produced by impacting drops of Newtonian fluids of various viscosities (mixtures of water and glycerol and a micellar system below the percolation threshold). In figure~\ref{fig:dversust}a we show the time evolution of the sheet diameter for samples with various zero-shear viscosity, $\eta_0$. For the sake of clarity, we only show data for glycerol/water mixtures, but data with micellar and microemulsion systems below the percolation threshold display similar features. As $\eta_0$ increases, the sheet expands less: the maximum diameter $d_{\rm{max}}$ is smaller and is reached earlier ($t_{\rm{max}}$ decreases). We quantify how the maximal expansion depends on viscosity and plot in figure~\ref{fig:dmax} the maximal diameter normalized by its value at small viscosity (the inviscid case), $\tilde{d}=\frac{d_{\rm{max}}}{d_{\rm{max}}(\eta_0 \rightarrow 0)}$ as a function of the zero-shear viscosity (bottom $x$-axis). We find that the maximal diameter is constant $d_{\rm{max}} = (21.2 \pm 1.2$) mm,  and independent of the viscosity for $\eta_0\leq \eta_c \sim 30$ mPa.s, and then continuously decreases with $\eta_0$. We notice that, in the viscosity regime where the sheet dynamics is affected by the viscosity ($\eta_0 > \eta_c$), volume loss due to ejection of secondary droplets is negligible.

We note that data obtained for pure water/glycerol mixture and for water/glycerol mixtures containing also surfactant and salt perfectly collapse over the whole range of viscosity investigated. The two classes of mixtures differ by their equilibrium surface tension: $\gamma = (66.9 \pm 3.4)$ mN.m$^{-1}$ for pure water/glycerol and $\gamma = (38.5 \pm 2.0)$ mN.m$^{-1}$ for the mixtures with CpCl and NaCl. Because of this difference the diameter of the needle used to dispense the drops has been modified in order to obtain droplets of equal size for all water/glycerol mixtures. However, the collapse of the two sets of data suggests that the equilibrium surface tension is not relevant to account for the sheet expansion. This is presumably due to the fact, that on the time scale needed to reach the maximal expansion (at most $4$ ms), the surface tension of the air/liquid interface does not have time to reach its equilibrium value. This time scale can be estimated as
$t_{\rm{d}} = \dfrac{\pi}{4 D}{\left( \dfrac{\Gamma_{\rm{eq}}}{C_{\rm{B}}}\right)} ^2$, where $D$ is the diffusion coefficient of the surfactant molecules, $\Gamma_{\rm{eq}}$ is the surface concentration of the surfactant at equilibrium and $C_{\rm{B}}$ is the bulk surfactant concentration \cite{Bonfillon1994}. For CpCl, $\Gamma_{\rm{eq}}=5\times10^{-10}$ $\rm{mol/cm^2}$, and $D=8.6\times10^{-6}$ $\rm{cm^2/s}$ in water ($\eta_0=1$ mPa s) \cite{Rillaerts1982}, yielding for a bulk concentration of $5.88$ mM, $t_{\rm{d}}=0.7$ ms in pure water. Hence, because $D$ is expected to scale as the inverse of the viscosity, the time scale needed to reach the equilibrium surface tension exceeds the time scale needed to reach the maximal expansion of the sheet for viscosity larger than $\sim 5$ mPa s. Consequently, for large viscosity the relevant surface tension of the water/glycerol/CpCl mixtures is expected to be that of water/glycerol mixtures, as observed experimentally. Note finally that surface tension gradients are presumably not relevant in our experiments. Indeed, because of Marangoni stresses, such gradients would cause perforation of the sheet~\cite{Vernay2015b,Rozhkov2010}, which is never observed in our experiments.

The normalized maximal diameter, $\tilde{d}$, is also plotted as a function of the Ohnsesorge number, $Oh$ (top $x$ axis, fig.~\ref{fig:dmax}). The Ohnesorge number is a dimensionless number that represents the ratio of internal viscosity dissipation to surface tension energy: $Oh=\frac{\eta_0}{\sqrt{\rho \gamma d_o}}$, where $\rho$ is the sample density, $\eta_0$ its zero-shear viscosity, $\gamma$ the surface tension and $d_0$ the drop diameter. Accordingly, for small $Oh$ ($Oh<\sim 0.1$), the maximal expansion of the sheet is governed by a balance between inertia and surface tension, whereas the higher $Oh$, the more dominant the viscous dissipation is. In the next section, we provide a modeling of the sheet expansion where inertia, surface tension and viscosity are taken into account.

\subsubsection{Viscoelastic samples}

For viscoelastic fluids, a relevant parameter is the Deborah number, $D_e$, defined as the ratio between the characteristic relaxation time of the viscoelastic samples and the lifetime of the sheet (typically $10$ ms). From the viscoelasticity measurements (fig.~\ref{fig:tau}a), one deduces that the only class of samples for which $D_e$ is significantly larger that $1$ is the microemulsions linked by $\rm{C_{18}}$ telechelic polymers. For the $\rm{C_{18}}$-based micelles, $D_e \cong 1$, whereas for the $\rm{C_{12}}$-based samples, $D_e \ll 1$.

We first focus on the maximal expansion of the sheet for $\rm{C_{12}}$-based micelles and microemulsions for which $De\ll1$. We find that these samples, although being viscoelastic, behave as purely viscous samples. This is shown qualitatively in the images displayed in figure~\ref{fig:imagesViscous}b taken at the maximal expansion of sheets produced with   $\rm{C_{12}}$-based micelles, and more quantitatively in figure~\ref{fig:dmax}. We indeed measure that the evolution of the normalized maximal expansion diameter, $\tilde{d}$, with the viscosity, $\eta_0$, for the $\rm{C_{12}}$-based micelles and microemulsions nicely superimpose over the whole range of viscosity investigated with the data acquired for water/glycerol mixtures.

By contrast, viscoelastic samples with characteristic times much larger than the life time of the sheet, as the $\rm{C_{18}}$-based microemulsions,  behave in a drastically different fashion. In the high Deborah number regime, we indeed measure that the sheets expand much more than viscous sheets or sheets produced with a viscoelastic samples with a short relaxation time. Remarkably, we find that up to zero-shear viscosity of $20$ Pa.s, the maximal expansion of the sheet does not significantly decrease compared to the case of inviscid fluids, whereas
for the other samples ($\rm{C_{12}}$-based micelles and microemulsions, and $\rm{C_{18}}$-based micelle), the maximum diameter of the sheet expansion is smaller than the target diameter ($\tilde{d} < 0.25$). We note moreover that quantitatively comparable results are obtained for regular microemulsions and for dyed ones.

Interestingly, the behavior of $\rm{C_{18}}$-based micelles for which the Deborah number is of the order of $1$ is intermediate between that of the $\rm{C_{12}}$-based ($D_e\ll1$), and that of the $\rm{C_{18}}$-based microemulsions ($D_e\gg1$), confirming a crucial role of the coupling between the characteristic relaxation time of the samples and the life time of the sheet.

\section {Discussion}
\label{Sec:discussion}

We provide here a modeling of the data obtained for Newtonian viscous  samples (micellar and micro-emulsions-based samples below the percolation threshold and glycerol/water mixtures) and also for viscoelastic fluids prepared with $\rm{C}_{12}$ telechelic polymer for which the Deborah number $De\ll1$. We  then discuss the interplay between viscosity and elasticity in the dynamics of viscoelastic sheets.

\subsection{Modeling of the maximal expansion of viscous sheets}

We consider here viscous samples and viscoelastic fluids based on $\rm{C}_{12}$ polymers. Those viscoelastic samples, for which $D_e\ll1$, are found to behave as viscous samples, the dependence of the maximal expansion of the sheet with the viscosity being equal to that of Newtonian fluids (fig.~\ref{fig:dmax}).

\subsubsection{Energy conservation}

We use as starting point a standard energy conservation approach. For an inviscid fluid, the kinetic energy of the drop upon impact is mainly converted into surface energy~\cite{Marmottant2000, Villermaux2011}. A rough estimate of the maximal expansion diameter of the sheet can be derived by balancing the kinetic energy of the drop and the surface energy at the maximal expansion of the sheet. In the case of a viscous drop, some energy is dissipated during the process, reducing the inertial expansion. We  assume that part of the impact energy is dissipated by a radial flow in the liquid sheet. Balancing the initial kinetic energy  against the surface energy at the maximal extension of the sheet and the viscous dissipation energy $E_{\rm{diss}} $ yields:

\begin{equation}
\label{eqn:energy1}
\frac{1}{2}mv_0^2 \approx \frac{\gamma \pi d_{\rm{max}}^2}{2} +E_{\rm{diss}}
\end{equation}

Here, $m$ is the mass of the drop, $v_{0}$ the impact velocity and $\gamma$ the surface tension.  Note that Eq.~\ref{eqn:energy1} neglects the interfacial energy between the target and the liquid (this is expected to hold as long as $d_{\rm{t}} \ll d_{\rm{max}}$)  and assumes that the sheet has a disk shape (the external rim is neglected). We define an effective velocity $v_{\rm{eff}}$ as the impact velocity of a fictive drop of the same but \textit{inviscid} fluid that will lead to the same maximal extension of the sheet. By definition no dissipation occurs in this case, and the energy balance reads:

\begin{equation}
\label{eqn:energy2}
\frac{1}{2}mv_{\rm{eff}}^2 \approx \frac{\gamma \pi d_{\rm{max}}^2}{2}
\end{equation}

Note that here we neglect the dissipation that might occur in the rim due to vortical flows~\cite{Clanet2004}. The dissipation energy therefore reads $E_{\rm{diss}}=\frac{1}{2}mv_0^2\big(1-\Big (\frac{v_{\rm{eff}}}{v_0}\Big)^2\big )$. The maximal diameter of the sheet normalized by its value in the inviscid case is obtained by considering Eq.~\ref{eqn:energy1} (with $E_{\rm{diss}}=0$) and Eq.~\ref{eqn:energy2} yielding:

\begin{equation}
\label{eqn:dnorm}
\tilde{d} \approx \frac{v_{\rm{eff}}}{v_0}
\end{equation}

\subsubsection{Theoretical approach}

To model the effective velocity, we make a parallel with experiments on so-called Savart sheets, formed through the impact of a slender jet on a small disc at high Reynolds number~\cite{Clanet2002}. In the stationary regime, the effect of the disc is to induce a shear viscous dissipation, which alters the liquid film velocity  at the edge of the target leading to~\cite{Yarin2006, Clanet2001, Roisman2009, Villermaux2013}:
\begin{equation}
\label{eqn:Ue}
 \frac{v_{\rm{eff}}}{v_0} \approx \frac{1}{1+\beta}
 \end{equation}

 Here $v_{\rm{eff}}$ is the velocity of the liquid film at the edge of the target, $v_0$ the impact velocity of the liquid jet, and $\beta$ is the ratio between the thickness of the viscous boundary layer, $\delta$, and the thickness of the fluid sheet at the edge of the target, $h$, as calculated from mass conservation. This model ensures that outside the target region, the flow of the  viscous sheet emerging from the impacting jet is identical to the flow of an inviscid liquid jet, provided that the  impact velocity $v_0$  is replaced by the effective viscosity $v_{\rm{eff}}$. On the other hand it has been shown theoretically~\cite{Villermaux2011} that the kinematics fields of the liquid sheet arising from the impact of a drop of an inviscid fluid onto a small target are  a time-dependent adaptation of a steady-state axisymmetric solution of Euler equations for a continuous jet impacting a solid target. These predictions have been confirmed experimentally~\cite{Vernay2015a, Vernay2015b}. Hence, following Refs.~\cite{Clanet2002,Villermaux2011} we assume that the effect of the viscous dissipation after the impact of the drop on the target with an impact velocity $v_0$ can be evaluated by considering that the flow of the  viscous sheet is equivalent to the flow of an \textit{inviscid} liquid sheet with an impact viscosity $v_{\rm{eff}}$ given by Eq.~\ref{eqn:Ue}. The scaling of the parameter $\beta$ is however different when one considers a jet and a drop impacting the target. In the case of a single drop of diameter $d_0$ hitting a target of diameter $d_{\rm{t}}$, simple mass conservation yields for the thickness of the fluid sheet when it fully covers the target (but does not yet expands in air):

\begin{equation}
\label{eqn:h}
h \approx \frac{2}{3}d_0 (d_0/d_{\rm{t}})^2
\end{equation}

On the other hand, the thickness of the viscous  boundary layer reads $\delta=\sqrt{\frac{\eta T}{\rho}}$  with $T \approx \frac{d_{\rm{t}}}{2v_0}$ the time needed to reach the edge of the target. Hence:

\begin{equation}
\label{eqn:delta}
\delta \approx \sqrt{\frac{\eta d_{\rm{t}}}{2 \rho v_0}}
\end{equation}

In the case of a single drop the adimensional parameter $\beta=\frac{\delta}{h}$ therefore reads:

\begin{equation}
\label{eqn:beta}
\beta \approx \frac{3}{2\sqrt{2}}\frac{1}{\sqrt{Re}} \big (\frac{d_{\rm{t}}}{d_0} \big )^{5/2}
\end{equation}

where $Re=(\rho v_0 d_0)/\eta$ is the Reynolds number. In our experimental conditions, $\beta$ varies been $0.03$ to $0.96$. From Eq.~\ref{eqn:energy2}, one therefore derives a simple expression for the  maximal extension of the sheet:

\begin{equation}
 \label{eqn:fit}
\frac{d_{\rm{max}}}{d_0} \approx \sqrt{\frac{We}{6}}\frac{1}{1+\beta}
\end{equation}

where $We=(\rho u_0^2 d_0)/\gamma$ is the Weber number, $\rho$ the mass density of the liquid and $\gamma$ the surface tension.

Once written for parameters normalized by their values in the inviscid case, Eq.~\ref{eqn:fit} reads:

\begin{equation}
 \label{eqn:fitNorm}
\tilde{d} \approx \frac{1}{1+\beta} \approx \frac{1}{1+\alpha \sqrt{\eta}}
\end{equation}

with

\begin{equation}
 \label{eqn:alpha}
\alpha=\frac{3}{2\sqrt{2}}\frac{1}{\sqrt{\rho v_0 d_0}} \big (\frac{d_{\rm{t}}}{d_0} \big )^{5/2}
\end{equation}

Equation~\ref{eqn:fitNorm} shows that the maximal extension continuously decreases with the square root of the sample viscosity. We check quantitatively this simple prediction with the experimental results obtained from water/glycerol mixtures and $\rm{C}_{12}$-based viscoelastic fluids of various viscosities (fig.~\ref{fig:dmax}). The dashed line (fig.~\ref{fig:dmax}) is the theoretical prediction (Eq.~\ref{eqn:fitNorm}) using the calculated value $\alpha=(1 \pm 0.1)$ $\rm{(Pa.s)}^{-0.5}$ (Eq.~\ref{eqn:alpha}). Error bars for $\alpha$ come from uncertainties on the drop and target diameters. Note that here we have taken an average value for the water/glycerol  density ($1125$ kg/m$^3$) and have neglected the increase of the density as the mixture gets enriched in glycerol, this increase (at most $20$ \%) being negligible compared to that of the viscosity ($3$ orders of magnitude). We find that the theoretical model provides a correct order of magnitude of the normalized diameter but systematically underestimates the dissipation. A fit of the experimental data using $\alpha$ as unique fitting parameter yields $\alpha=(1.46\pm0.13)$ $\rm{(Pa.s)}^{-0.5}$  but does not provide a significantly better account of the viscosity-dependence of the normalized maximal diameter ({\color {myc} dotted line in fig.~\ref{fig:dmax}}).

\subsubsection{Semi-empirical approach}

\begin{figure}
\includegraphics[width=0.5\textwidth]{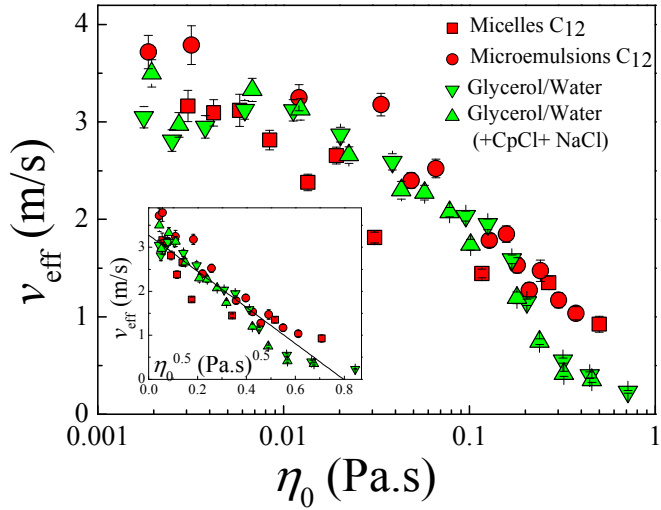}
\caption{(Color online) Effective velocity of the sheet measured at the edge of the target, $v_{\rm{eff}}$ , as a function of the zero-shear viscosity $\eta_0$, for the viscous samples and the viscoelastic samples for which $De \ll 1$. (Main plot) lin-log scale, and (inset) same data plotted as a function of $\sqrt{\eta_0}$ in a lin-lin scale.}
\label{fig:speed}
\end{figure}

To better account for the viscosity dependence of the maximal expansion, we still use a similar approach based on energy conservation and consider that a viscous sheet expands as an inviscid one but with an effective impact velocity $v_{\rm{eff}}$ reduced compared to the real one. This velocity is assumed to be the velocity at the edge of the target and is reduced compared to the initial velocity of the sheet due to viscous dissipation on the surface of the target. Instead of using a theoretical model for $v_{\rm{eff}}$ as above (Eq.~\ref{eqn:Ue}), we here measure directly $v_{\rm{eff}}$ as the time derivative at short time of $d$ the diameter of the sheet: $v_{\rm{eff}}=\frac{1}{2}\frac{\partial d}{\partial t} )_{t \rightarrow 0}$. We show in figure~\ref{fig:speed} $v_{\rm{eff}}$ for all the viscous samples and viscoelastic samples with $De \ll 1$.  Data for all samples reasonably collapse and show a continuous decrease with the sample viscosity. We find that a linear dependence of $v_{\rm{eff}}$ with
$\sqrt{\eta_0}$ accounts fairly well for the dependence of $v_{\rm{eff}}$ with the sample viscosity (inset fig.~\ref{fig:speed}). Note that other simple functional forms, as for instance a linear dependence with the viscosity or the theoretical expectation (Eq.~\ref{eqn:Ue}), are significantly less good. A fit of the experimental data with the functional form $v_{\rm{eff}}=v^* (1-\lambda \sqrt{\eta_0})$ yields $v^*=(3.2 \pm 0.1)$ m/s and $\lambda=1.24  \pm 0.10$ $\rm{(Pa.s)}^{-0.5}$. Note that $v^*$ is $20$ \% smaller than the impact velocity ($v_0=4$ m/s), indicating that some dissipation also occurs in the inviscid case. From Eq.~\ref{eqn:dnorm} one therefore predicts:

\begin{equation}
 \label{eqn:semiempirical}
\tilde{d}=v_{\rm{eff}}/v^*=1-\lambda \sqrt{\eta_0}
\end{equation}

{\color {myc} We plot in figure~\ref{fig:dmax}  Eq.~\ref{eqn:semiempirical} with the numerical value of $\lambda$ extracted from the fit of the effective velocity (fig.~\ref{fig:speed}). In spite of its simplicity, we find that the semi-empirical approach provides a good agreement of the experimental data, especially in the regime of relatively high viscosity, where the theoretic model fails}.
We note also that $1-\lambda \sqrt{\eta_0}$ in Eq.(\ref{eqn:semiempirical}) is the limit of $1/(1+\lambda\sqrt{\eta_0})$ for $\lambda^2\eta_0\ll1$, a functional form that corresponds to the theoretical model (Eq.~\ref{eqn:fitNorm}).

\begin{figure}
\includegraphics[width=0.5\textwidth]{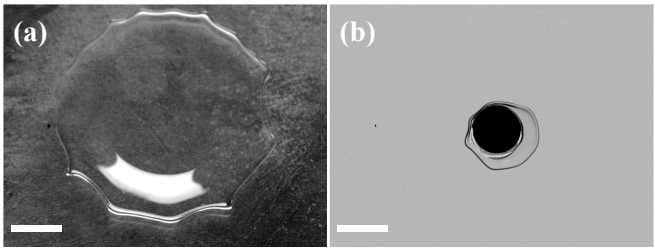}
\caption{Images on viscous sheet taken at the maximal expansion for a drop of water/glycerol mixture with zero-shear viscosity $227$ mPa.s impacting at the same velocity (a) a silicon wafer covered with a thin layer of liquid nitrogen and (b) a small solid target. Scale bars: $6$ mm.}
\label{fig:nitrogen}
\end{figure}

To confirm the crucial role played by the viscous dissipation on the surface of the small target, we have performed an experiment where we expect to suppress any dissipation on a solid surface. To do so, a drop is impacted (with the same velocity as for previous experiments on a solid target) on a surface covered by a thin layer of liquid nitrogen, allowing the drop and the sheet to float above a gaseous cushion. Comparative images of the sheets at maximal expansion are shown in figure~\ref{fig:nitrogen}: for a same mixture of water and glycerol with a zero-shear viscosity of $227$ mPa.s, $d_{\rm{max}}= 10.0$ mm when the drop is impacting on the small solid target, whereas $d_{\rm{max}}$ reaches $25.1$ mm on nitrogen. This expansion is comparable to that of a fluid with a very low Newtonian viscosity, demonstrating that viscous dissipation on the solid target governs the expansion dynamics.

\subsection{Viscoelastic sheets}

\begin{figure}
\includegraphics[width=0.5\textwidth]{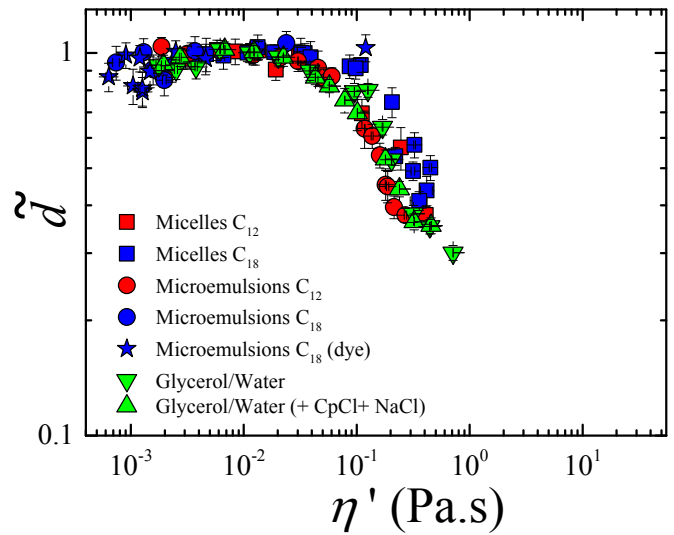}
\caption{(Color online) Same data as in figure~\ref{fig:dmax} but plotted as a function of the dynamic viscosity.}
\label{fig:complexviscosity}
\end{figure}

Our experimental observations show, for samples with zero-shear viscosity larger than $\sim 30$ mPa.s, a clear link between the capability of a sheet to expand and the linear rheological properties of the fluid.
A simple yet quantitative way to account for the samples viscoelasticity is to consider the dynamic viscosity. Note that this approach is similar to that followed by de Gennes to describe the fracture dissipation process for complex fluids in the framework of the trumpet model~\cite{Tabuteau2011, PGG1988, PGG1996, Saulnier2004}.
For a Maxwell fluid, the dynamic viscosity is $\eta'=G"/\omega=\frac{G_0 \tau}{1+(\omega \tau)^2}$. Taking as characteristic relaxation time, $t_{\rm{max}}$, the time at which the sheet reaches its maximal expansion, $\eta'=\frac{\eta_0 }{1+D_e^2}$, with $D_e=\tau/t_{\rm{max}}$. Interestingly, if the data of the maximal expansion are plotted as a function of $\eta '$ instead of $\eta_0$ we find that the data for $\rm{C_{18}}$-based micelles (for which $D_e \approx 1$) and $\rm{C_{18}}$-based microemulsions (for which $D_e \gg1$) almost collapse with the data gathered for Newtonian fluids and $\rm{C_{12}}$-based samples (fig.~\ref{fig:complexviscosity}). This suggests that the deviation from the Newtonian behavior, which is observed for viscoelastic samples whose relaxation time is comparable to, or larger than, the life time of the sheet, can be accounted for by the sample linear viscoelasticity.

\begin{figure}
\includegraphics[width=0.5\textwidth]{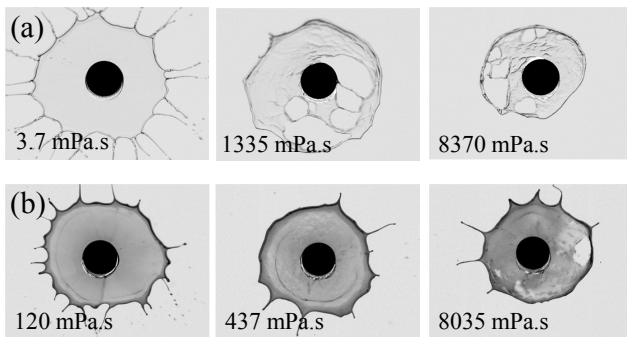}
\caption{Snapshots of sheets made of $\rm{C_{18}}$-based microemulsions (a) without dye, (b) with dye. Images are taken at the maximal expansion of the sheets. All samples are above the percolation threshold, except the one with the lower zero-shear viscosity ($3.7$ mPa.s).}
\label{fig:imagesElastic}
\end{figure}

For $\rm{C_{18}}$-based microemulsions, the characteristic relaxation time being much larger than the life time of the sheet ($D_e \gg 1$), the dynamic viscosity is much smaller than the zero-shear viscosity, indicating that dissipation is negligible, and the samples behave as almost purely elastic. Based on previous experiments in different experimental conditions, we know that, for those kinds of samples, fracture processes occur: cracks have indeed been observed  in the gap of a shear cell (pure shear flow~\cite{Tabuteau2009}), in a pendant drop experiment (pure extensional flow~\cite{Tabuteau2009}) and in Hele-Shaw cell (complex flow involving both shear and extensional flows~\cite{Mora2010, Foyart2013}). Fracture occurs when the sample is deformed at rates larger than the inverse of its viscoelastic relaxation time.  It is therefore instructive to evaluate the  deformation rates involved upon the sheet expansion. The extension rate of the fluid upon the expansion of the sheet, $\dot{\epsilon}$, can be computed as $\dot{\epsilon}=\frac{1}{d}\frac{\partial d}{\partial t}$. Note that at short times, $\dot{\epsilon}=\frac{2}{d_{\rm{t}}}v_{\rm{eff}}$, with $d_{\rm{t}}=6$ mm the target diameter, and $v_{\rm{eff}}$ the effective velocity determined above (fig.~\ref{fig:speed}), yielding a rate $\dot{\epsilon} \approx 1000 \, \rm{s}^{-1}$ at short time that steadily decreases with time until vanishing when the sheet reaches its maximal expansion. We can also evaluate the shear rate involved as the drop hits the target. A simple estimation, by considering that shear flow only occurs when the fluid interacts with the solid surface, gives $\dot{\gamma}\approx\frac{v_{\rm{eff}}}{\delta}$, with $\delta$ the thickness over which viscous dissipation takes place (Eq.~\ref{eqn:delta}). For $\rm{C_{18}}$-based microemulsions above percolation, $v_{\rm{eff}}\approx3.5$ m/s, and $\dot{\gamma}$ is in the range $(1000-10000)$ $\rm{s}^{-1}$. Hence, the rates involved are systematically larger than the inverse of the relaxation time ($\tau$ is in the range $(0.1-1)$ s, fig.~\ref{fig:tau}a), suggesting that fracture processes may occur.  Consistently, the sheets produced with the $\rm{C_{18}}$-based microemulsion samples exhibit systematically a very anomalous aspect. As shown for both dyed and undyed samples (fig.~\ref{fig:imagesElastic}), cracks and holes invade the interior of the sheets, while the sheet maintains the integrity of its contour. Hence, it is tempting to associate the fracture process that occurs in extension and shear for $\rm{C_{18}}$-based microemulsion samples to the cracks observed when the sheet expands. This would require a theoretical rationalization of the nucleation and instability processes as the sheet expands, which is out of the scope of the manuscript. Our observations nevertheless clearly emphasize that the elastic contribution in the viscoelasticity plays a crucial role in the overall way a sheet expands and are reminiscent of the rupture processes of Savart sheets observed upon collision of viscoelastic jets~\cite{Miller2005}. Note that, in sharp contrast, in the low number Deborah regime, in which the viscoelastic sheet behaves as viscous sheets, the sheet expands in a rather smooth way and always preserves its integrity (fig.~\ref{fig:imagesViscous}).

\section {Conclusion}

We have investigated how viscous and viscoelastic sheets resulting from the impact of a droplet on a small target of size comparable to that of the droplets expand in air. We have found that, for viscous droplets and for small Ohnesorge number $Oh$, the maximal expansion of the sheet is governed by inertia and surface tension, whereas for higher $Oh$ viscous dissipation plays an important role. We have used  a simple modeling of the sheet expansion where inertia, surface tension and viscosity are taken into account to compute an energy balance. One of the main ingredients is that dissipation occurs in the viscous boundary layer at the interface between the surface of the target and the liquid. Accordingly, we have directly measured the sheet velocity at the edge of the target and relate this velocity to the maximal expansion of the sheet. The quantitative agreement between the model and the experiments shows that the dissipation mainly occurs on the small surface of the target and that, when the sheet freely expands in air, dissipation is negligible, in agreement with an experiment performed with a viscous sample freely expanding on a gaseous nitrogen cushion.

In addition, we have investigated model complex fluids that behave as Maxwell fluids, characterized by a plateau modulus $G_0$ and a unique relaxation time $\tau$. We have been able to independently and finely tune $\tau$ and the zero-shear viscosity $G_0 \tau$, allowing us to decouple the effect of the viscosity and elasticity on the spreading of viscoelastic sheets. Our experiments have evidenced a key role of the adimensional Deborah $D_e$ number defined as the ratio between $\tau$ and the typical lifetime of the sheet. When $D_e\ll1$, we find that the Maxwell fluids behave as simple Newtonian fluids with zero-shear viscosity $G_0 \tau$. For samples for which $D_e > 1$, the departure from the Newtonian case can be accounted for by the sample viscoelasticity. Interestingly, the dynamics of viscoelastic sheets produced with Maxwell fluids whose characteristic relaxation time is much larger than the life time of the sheet ($D_e\gg1$) drastically differ from those of other samples. The sheet expands much more than viscous sheets with comparable zero shear viscosity, due to reduced viscous dissipation. In addition, upon expansion the sheet is highly heterogeneous and displays cracks and holes, revealing the sample elasticity.
Our results therefore show  the interplay between the relaxation time of the viscoelastic fluid and the characteristic time of the experiments, as could also be observed in other experimental conditions like the dispensing of drop from a syringe with a controlled flow rate~\cite{Clasen2012}.
The experimental setup, originally designed by M. Vignes-Adler and collaborators and subsequently modified by Villermaux and Bossa to limit the bell effect, provides moreover an unusual yet interesting configuration for testing the rheology of complex fluids as it involves large deformations and highly extensional flows on very short time scales.

\begin{acknowledgments}
 This work has been supported by the E.U. (Marie Sklodowska-Curie ITN "Supolen", no. 607937). We thank Serge Mora and Henri Lhuissier for fruitful discussions, and Ty Phou and Jean-Marc Fromental for technical assistance.

\end{acknowledgments}



\begin{thebibliography}{26}
\expandafter\ifx\csname natexlab\endcsname\relax\def\natexlab#1{#1}\fi
\expandafter\ifx\csname bibnamefont\endcsname\relax
  \def\bibnamefont#1{#1}\fi
\expandafter\ifx\csname bibfnamefont\endcsname\relax
  \def\bibfnamefont#1{#1}\fi
\expandafter\ifx\csname citenamefont\endcsname\relax
  \def\citenamefont#1{#1}\fi
\expandafter\ifx\csname url\endcsname\relax
  \def\url#1{\texttt{#1}}\fi
\expandafter\ifx\csname urlprefix\endcsname\relax\def\urlprefix{URL }\fi
\providecommand{\bibinfo}[2]{#2}
\providecommand{\eprint}[2][]{\url{#2}}


\bibitem{Yarin2006} A. L. Yarin, Ann. Rev. Fluid Mech., \textbf{38}, 159, 2006.

\bibitem{Josserand2016} C. Josserand and S. T. Thoroddsen, Ann. Rev. Fluid Mech., \textbf{48}, 365, 2016.

\bibitem{German2009a} G. German and V. Bertola, Atomization Sprays, \textbf{19}, 787, 2009.

\bibitem{Scheller1995} B. L. Scheller and D. W. Bousfield, AIChE J., \textbf{41}, 1357, 1995.

\bibitem{Bertola2013} V. Bertola, Adv. Colloid Interface Sci., \textbf{193}, 1, 2013.

\bibitem{Cooper-White2002} J. J. Cooper-White, R. C. Crooks and D. V. Boger, Colloids Surfaces A, \textbf{210}, 105, 2002.

\bibitem{Nigen2005} S. Nigen, Atomization Sprays, \textbf{15}, 103, 2005.

\bibitem{German2009b} G. German and V. Bertola, J. Phys.: Condens. Matter, \textbf{21}, 375111, 2009.

\bibitem{Luu2009} L.-H. Luu and Y. Forterre, J. Fluid Mech., \textbf{632}, 301, 2009.

\bibitem{Luu2013} L.-H. Luu and Y. Forterre, Phys. Rev. Lett., \textbf{110}, 184501, 2013.

\bibitem{Saidi2010} A. Sa\"{i}di, C. Martin and A. Magnin, J. Non-Newtonian Fluid Mech.,  \textbf{165}, 596, 2010.

\bibitem{Nicolas2005} M. Nicolas, J. Fluid Mech., \textbf{545}, 271, 2005.

\bibitem{Laan2014} N. Laan, K. G. de Bruin, D. Bartolo, C. Josserand and D. Bonn, Phys. Rev. Appl., \textbf{2}, 044018, 2014.

\bibitem{Bartolo2007} D. Bartolo, A. Boudaoud, G. Narcy and D. Bonn, Phys. Rev. Lett., \textbf{99}, 174502, 2007.

\bibitem{Bertola2015} V. Bertola and M. Wang, Colloids Surfaces A, \textbf{481}, 600, 2015.

\bibitem{Bolleddula2010} D. A. Bolleddula, A. Berchielli and A. Aliseda, Adv. Colloid Interface Sci., \textbf{159}, 144, 2010.

\bibitem{Mao1997} T. Mao, D. C. S. Kuhn and H. Tran, AIChE J., \textbf{43}, 2169, 1997.

\bibitem{Rozhkov2002} A. Rozhkov, B. Prunet-Foch and M. Vignes-Adler, Physics Fluid,  \textbf{14}, 3485, 2002.

\bibitem{Rozhkov2004} A. Rozhkov, B. Prunet-Foch and M. Vignes-Adler, Proc. Royal Soc. A,  \textbf{460}, 2681, 2004.

\bibitem{Villermaux2011}  E. Villermaux and B. Bossa, J. Fluid Mech. \textbf{668}, 412, 2011.

\bibitem{Vernay2015a} C. Vernay, L. Ramos and C. Ligoure, J. Fluid Mech., \textbf{764}, 428, 2015.

\bibitem{Vernay2015b} C. Vernay, L. Ramos and C. Ligoure, Phys. Rev. Lett., \textbf{115}, 198302, 2015.

\bibitem{Savart1833} F. Savart, Ann. Chim. Phys., \textbf{LIX}, 266, 1833.

\bibitem{Rozhkov2006} A. Rozhkov, B. Prunet-Foch and M. Vignes-Adler, J. Non-Newtonian Fluid Mech.,  \textbf{134}, 44, 2006.

\bibitem{Jung2011} S. Jung, S. D. Hoath, G. D. Martin and I. M. Hutchings, J. Non-Newtonian Fluid Mech.,  \textbf{166}, 297, 2011.

\bibitem{Rozhkov2015} A. Rozhkov, B. Prunet-Foch and M. Vignes-Adler,  J. Non-Newtonian Fluid Mech.,  \textbf{226}, 46, 2015.

\bibitem{Ligoure2013} C. Ligoure and S. Mora, Rheol. Acta, \textbf{52}, 91, 2013.

\bibitem{Tixier2010} T. Tixier, H. Tabuteau, A. Carri\`{e}re, L. Ramos and C. Ligoure, Soft Matter, \textbf{6}, 2699, 2010.

\bibitem{Michel2000} E. Michel, M. Filali, R. Aznar, G. Porte and J. Appell, Langmuir, \textbf{16}, 8702, 2000.

\bibitem{Tabuteau2009} H. Tabuteau, S. Mora, G. Porte, M. Abkarian and C. Ligoure, Phys. Rev. Lett., \textbf{102}, 155501, 2009.

\bibitem{Tabuteau2011} H. Tabuteau, S. Mora, M. Ciccotti, C. Y. Hui and C. Ligoure, Soft Matter, \textbf{7}, 9474, 2011.

\bibitem{Bonfillon1994} A. Bonfillon, F. Sicoli and D. Langevin, J. Colloid Interface Sci., \textbf{168}, 497, 1994.

\bibitem{Rillaerts1982} E. Rillaerts and P. Joos, J. Phys. Chem., \textbf{86}, 3471, 1082.

\bibitem{Rozhkov2010} A. Rozhkov, B. Prunet-Foch and M. Vignes-Adler, Proc. Royal Soc. A,  \textbf{466}, 2897, 2010.

\bibitem{Marmottant2000}  P. Marmottant, E. Villermaux and C. Clanet, J. Colloid Interface Sci., \textbf{230}, 29, 2000.

\bibitem{Clanet2004} C. Clanet, C. B\'{e}guin, D. Richard and D. Qu\'{e}r\'{e}, J. Fluid Mech. \textbf{517}, 199, 2004.

\bibitem{Clanet2002} C. Clanet and E. Villermaux, J. Fluid Mech. \textbf{462}, 307, 2002.

\bibitem{Clanet2001} C. Clanet, J. Fluid Mech. \textbf{430}, 111, 2001.

\bibitem{Roisman2009} I. V. Roisman, Phys. Fluids, \textbf{21}, 052104, 2009.

\bibitem{Villermaux2013}  E. Villermaux, V. Pistre and H. Lhuissier, J. Fluid Mech. \textbf{730}, 607, 2013.

\bibitem{PGG1988} P.-G. de Gennes, CR Acad. Sci. Paris Serie II, \textbf{307}, 1949, 1988.

\bibitem{PGG1996} P.-G. de Gennes, Langmuir, \textbf{12}, 4497, 1996.

\bibitem{Saulnier2004} F. Saulnier, T. Ondar\c{c}uhu, A. Aradian and E. Rapha\"{e}l, Macromolecules \textbf{37}, 1067, 1996.

\bibitem{Mora2010} S. Mora and M. Manna, Phys. Rev. E, \textbf{81}, 026305, 2010.

\bibitem{Foyart2013} G. Foyart, L. Ramos, S. Mora and C. Ligoure, Soft Matter, \textbf{9}, 7775, 2013.

\bibitem{Miller2005} E. Miller, B. Gibson, E. McWilliams and J. P. Rothstein, Applied Physics Lett., \textbf{87}, 014101, 2005.

\bibitem{Clasen2012} C. Clasen, P. M. Phillips, L. Palangetic and J. Vermant, AIChE J., \textbf{58}, 3242, 2012.



\end{thebibliography}

\end{document}